\documentclass[aps,pre,showpacs]{revtex4}
\usepackage{graphicx}
\usepackage{amssymb}
\begin{document}
\title{Dipole fluid as a basic model for the equation of
state of ionic liquid \\ in the vicinity of their critical point}
\author{V.L. Kulinskii}
\email{koul@paco.net}
\author{N.P. Malomuzh}
\email{mnp@ukr.net} \affiliation{Department of Theoretical
Physics, Odessa National University, 2 Dvoryanskaya Str., 270100
Odessa, Ukraine}
\begin{abstract}
The model of dipole fluid for the ionic liquids similar to the
molten NaCl is proposed. The estimates for the critical parameters
are obtained with the help of the van der Waals equation of state.
The influence of the rotation on the characteristics of a dipole
pair and the location of the critical point is discussed. The
dissociation of such fluid near the critical point is considered.
\end{abstract}
 \maketitle

\section*{Introduction}

There is the strong indication for existence of the liquid-gas
critical point in ionic liquids like molten NaCl. The main model
for such systems is called Primitive Model or its restricted
version where the characteristics (mass, diameter etc.) of the
opposite charges are the same. Because of formation of associated
ionic pairs the derivation of mean field equation of state in
rigorous thermodynamic approach is yet unsolved. The existence of
additional constraints, so called sum rules \cite{mart}, makes the
problem of finding thermodynamically consistent mean field theory
hard to solve comparing with the molecular liquid case. The
various approaches have been proposed recently for the derivation
of equation of state (EOS) for the RPM. Mostly they are based
either on the improvements of classical Debye-H\"uckel (DH)
\cite{dh} and Bjerrum \cite{bj} theory of association or Mean
Spherical Approximation (MSA) \cite{msa} taking into account the
association and various effective interactions. For the reviews of
the results see \cite{fish,fishlev,st2,gg,ws}.

The location of the critical point for RPM varies in dependence of
the interactions included into free energy. It is accepted that
the following dimensionless parameters for the temperature $T$,
particle number density $n$ and pressure $P$ are used:
\begin{equation}\label{dim}
  T^{*}=\frac{k_{B}T}{q^2/a}\;,\quad \rho^{*}= n a^3\;,\quad
  P^{*}=\frac{P}{q^2/a^4}
\end{equation}
where $a$ is the ionic diameter $q$ is the absolute value of the
charge. The characteristic value for $T^{*}_{c}$ takes the values
in the interval $0.05 \le T^{*}_{c}\le 0.08 $. The situation with
the critical density $\rho_{c}$ is less definite. Its value varies
from $\rho_{c} = 0.02$ to $\rho_{c} = 0.08$. It should be noted
that most of the analytical results predict low density
($\rho_{c}=0.026$) and "high temperature" ($T^{*}_{c}= 0.06 -
0.08$) critical point. It cannot be excluded that such low value
for $\rho^{*}_{c}$ is connected with small association taken into
account with the help of thermodynamic perturbation theory.
However the reliability of such type estimates is not quite clear.
Therefore the development of the alternative approach grounded on
the dipole fluid model seem to be expedient.

The Monte Carlo (MC) simulation studies \cite{champ,mc,gr}
performed in recent years give "high density" $\rho_{c}=0.07 -
0.08$ and low temperature $T^{*}_{c}= 0.04 - 0.05$ critical point.

In the present paper we estimate the critical characteristics of
the ionic molten in the framework of the dipole fluid approach. We
will also show that the change of molecular parameters due to
rotations influences essentially the location of the critical
point.

\section{Qualitative analysis of the critical behavior
of the dipole liquid}

NaCl is the simplest example of ionic liquid. In solid state it is
ionic crystal. Above the melting point the positions of ions
become unfixed, but this liquid remains strongly dissociated. Due
to thermal expansion at increasing temperature the dissociation
degree diminishes and the molten salt passes to a dipole liquid.
At further increasing temperature and decreasing the molten
density the dissociation degree grows again and the molten salt
becomes completely ionized system. Thus, at some temperature
$(T_{1},T_{2})$ and density $(n_{1}, n_{2})$ intervals the molten
NaCl can be considered as a dipole liquid. Further the additional
arguments for this assumption will be given.

Let us consider the general properties of this liquid and, in
particular, its main critical parameters: the critical temperature
and density. The interparticle interaction in dipole system takes
the form:
\begin{equation} U(1,2)=U_{w}(1,2)+U_{dd}(1,2) \end{equation}
where the first term describes the Van-der-Waals interaction
between molecules and
\begin{equation} U_{dd}=\frac{1}{\epsilon
r_{12}^{3}}\left(\vec{d_{1}}\vec{d_{2}}-3\frac
{(\vec{d_{1}}\vec{r_{12}})(\vec{d_{2}}\vec{r_{12}})}{r_{12}^{2}}\right)
\end{equation} the proper dipole-dipole interaction,
$r_{12}=|\mathbf{r}_{1}-\mathbf{r}_{2}|$ is the interparticle
spacing. Note that the dipole moment of a pair is some function of
the equilibrium distance between ions in a pair:
\[\mathbf{d}_{i}=d(r_{d})\mathbf{n}_{i},\quad \mathbf{n}_{+}=
\frac{\mathbf{r}^{(i)}_{+}-\mathbf{r}^{(i)}_{-}}{|\mathbf{r}^{(i)}_{+}-\mathbf{r}^{(i)}_{-}|},\]
$i=1,2$, $\epsilon$ is dielectric permittivity.

Because of the dipole-dipole interactions are relatively weak the
angular distribution of dipole moments $\vec{d_{i}}$ is close to
isotropic one. More exactly we assume that the two particle
distribution function $g(\vec{d_{1}},\vec{d_{2}})$ can be
approximated by the first two terms in the expansion:
\begin{equation} g(\vec{d_{1}},\vec{d_{2}})=1-\beta U_{dd}(1,2)+...,
\qquad \beta=\frac{1} {k_{B}T}. \end{equation}
The approximation
of such a kind allows us to exclude the orientational degrees of
freedom in the configurational integral with the help of the
perturbation theory. In fact this procedure is equivalent to the
usage of the isotropic potential
\begin{equation} \label{dip}
\begin{array}{ll}
U(r_{12})=<U(1,2)>=U_{w}(r_{12})-U_{d}
\left(\frac{\sigma}{r_{12}}\right)^{6}&,\\
U_{d}=\frac{2}{3}\beta\frac{<<d^{2}>>^2} {\epsilon^{2}\sigma^{6}}&
\end{array}
\end{equation}
Here $\sigma \approx a_{+}+a_{-}\approx 2 a$, $a_{+}$ and $a_{-}$
are the diameters of ions Na and Cl correspondingly and for
simplicity we put $a_{+}=a_{-}$ and neglect the difference in
masses of the ions, $<<...>>$ denotes the average with internal
partition function of a pair. It is easy to check that the
inequality
$|U_{w}(r_{12})|<<U_{d}\left(\frac{\sigma}{r_{12}}\right)^{6} $
takes place at all $r_{12}$. Therefore further the contribution
$|U_{w}|$ will be ignored.

It is essential, that the averaging procedure restricts the
applicability region of the potential (\ref{dip}) by interparticle
spacings $\sigma \leq r_{12}$ which gives the size of the
"averaged" dipole of order $2a$. Though the value of $\sigma$ may
be slightly less than $2 a$ since rotating dipoles are not the
same as hard spheres of diameter $2a$. It is quite clear in view
of scattering cross section for the hard rotating dumbells. At
this level $\sigma$ should be considered as the parameter (in
general temperature dependent) of the dipole-dipole potential. The
procedure of its fixing in the critical point will be discussed
further.

To describe the properties of the molten NaCl within the interval,
where it can be considered as a dipole liquid, we can use the
potential with hard wall:

\begin{equation}
\label{ddp} U(r_{12})=\left \{ \begin{array}{ll} \infty &,\;
r_{12}< \sigma \\ -U_{d}\left(\frac{\sigma}{r_{12}}\right)^{6}&,\;
 \sigma \leq r_{12}
 \end{array}
 \right .
\end{equation}

Such a potential leads to the Van-der-Waals equation of state
\begin{equation}
\label{vdw} P=\frac{n_{d}k_{B}T}{1-n_{d} b}-A(T)n_{d}^{2},
\end{equation}
where
\begin{equation}
\label{A} A(T)=-\pi \int\limits_{\sigma}^{\infty} U(r,T)r^2 dr =
\frac{\pi \sigma^3}{3}U_{d} \;,\quad b=\frac{2\pi}{3}\sigma^3
\end{equation}
and $n_{d}$ is the pair number density. Therefore the overall
density is $n=2n_{d}$. In dimensionless form (\ref{vdw}) and
(\ref{A}) read as:
\begin{equation}
\label{vdwd} P^{*}=\frac{\rho^{*} T^{*}}{2 - b
\rho^{*}}-\frac{\tilde A(T^{*})}{4}\rho^{*^2},
\end{equation}
and
\begin{equation}
\label{tA} \tilde A(T^{*})=\frac{2\pi}{9
T^{*}\sigma^3}<<r_{d}^2>>^2
\end{equation}
Here all spatial parameters is given in units of $a$. The value of
parameter $<<r_d>>$ depends on the internal structure of the pair.
Though by the order of magnitude $<<r_d>>\approx 1$, nevertheless
from \cite{dip} it follows that the characteristic values of
dipole moments correspond to $<<r_d>>\, < \, 1$.

Since $\sigma$ is connected with the size of the pair we model its
temperature dependence via relation:
\begin{equation}\label{sig}
  \sigma =  <<r_{d}>> \delta
\end{equation}
where $\delta$ is the fitting parameter.

Note that $\sigma$ is temperature dependent which we assume the
same as that for $<<r>>$. The functions $<<r^n>>$ will be
determined below. Note that the vapor phase contacting with the
liquid one is the gas of dipole molecules. The van der Waals EOS
is appropriate approximation for EOS  for such vapor phase. Thus
we can get the the critical parameters of this system using the
van der Waals theory of the critical point.

The equation (\ref{vdw}) leads to the following equations for the
critical temperature and density (note that $n_{d}=n/2$, where $n$
is the total number density):
\begin{equation}
\label{prm} T^{*}_{c}=\frac{2\sqrt{2}}{9\sigma^3}<<r_{d}^2>>,
\qquad \rho^{*}_{c}=\frac{1}{\pi\sigma^3},
\end{equation}
The estimates for these parameters are straightforward if we put
$\sigma = 2$, and take into account that due to small dipole
moment of NaCl $<<r_{d}^2>>=1$ (in units of $a$):
\begin{equation}
\label{critest} T^{*}_{c}=\frac{\sqrt{2}}{36}\approx 0.04\;,
\qquad \rho^{*}_{c}=\frac{1}{8\pi}\approx 0.04\;,\quad
P^{*}_{c}=\frac{\sqrt{2}}{1536 \pi}\approx 3\cdot 10^{-4}
\end{equation}
\[Z_{c}=\frac{P^{*}_{c}}{\rho^{*}_{c}T^{*}_{c}}=\frac{3}{16}\approx 0.19\]
which are in satisfactory agreement with the values
\begin{equation}
\label{fl} T^{*}_{c}= 0.055\;, \qquad \rho^{*}_{c}= 0.026\;\quad
P^{*}_{c}=3.6\cdot 10^{-4},
\end{equation}
\[Z_{c}= 0.25\]
obtained within extended Debye-H\"uckel-Bjerrum theory
\cite{fishlev} augmented with ion-dipole interaction. Our value of
the  critical density is greater due to the neglecting the
dissociation of the dipole pairs.

Now we need to consider the dipole pair as itself since the
parameters of the potential (\ref{ddp}) actually are the averages
over the internal partition function of a pair and therefore are
the temperature dependent functions.

\section{The
dissociation of the rotating dipole liquid}

In previous section the model of undissociated ionic liquid
consisting of rotating dipoles has been introduced. Here we investigate
the internal structure of the bound pair of ions. We take into
account the fact that the energy of interaction of a pair should
include centrifugal energy together with Coulombic
potential as in standard problem of two
bodies interacting via central field.

The dissociation temperature for NaCl-molten is determined by the
effective potential of an ion within a rotating dipole which
includes the centrifugal energy:
\begin{equation}
\label{rrf} kT_{d}\approx -U_{eff}= \frac{q^2}{r}-\frac{L^2}{2I},
\end{equation}
where $I=\mu r^2$ is the moment of inertia of the charge
with reduced mass
$\mu=\frac{m_{+}m_{-}}{m_{+}+m_{-}}= m/2$.
At such high
temperatures all degrees of freedom are all in equilibrium and we
can use the estimate
\begin{equation}
\label{rotat} \label{rot} E_{rot}=\left<\frac{L^2}{2I}\right>=kT.
\end{equation} Note that the equilibrium distance between ions in
a pair, which is determined by the minimum of effective potential
(\ref{rrf}) with the help of (\ref{rotat}), is
\begin{equation}\label{bjer}
a_{eq}=\frac{1}{2T^{*}},
\end{equation}
which is exactly the Bjerrum size of the pair $R^{Bj}$ \cite{bj}
(see also \cite{fishlev}). The choice of (\ref{bjer}) as the size
of the ionic pair is inappropriate from the physical point of view
at low temperatures $T^{*}<<1$ \cite{fishlev}. It is natural that
with lowering $T$ the size of a pair should become smaller tending
to $a$ at $T\rightarrow 0$. That is why it was suggested to use it
for $1/T^{*}\ge 2$ only.

Let us consider this question within the picture formulated above.
To be more correct we will include the rotational energy into
association constant, which is proportional to the internal
partition function of the pair \cite{fishlev,gg}:
\begin{equation}
\label{as} K(T^{*},R)=4\pi\int\limits_{a}^{R}\exp(-\beta
U_{eff})r^2 dr
\end{equation}
In 2D case one can put $R=\infty$ because of the logarithmic
growth of the electrostatic potential and get the estimation of
Kosterlitz-Thouless temperature of dissociation \cite{kt}. In 3D
case there is the problem with upper cutoff in such an approach
where the association constant is identified with internal
partition function of the ionic pair.

\begin{figure}
\includegraphics{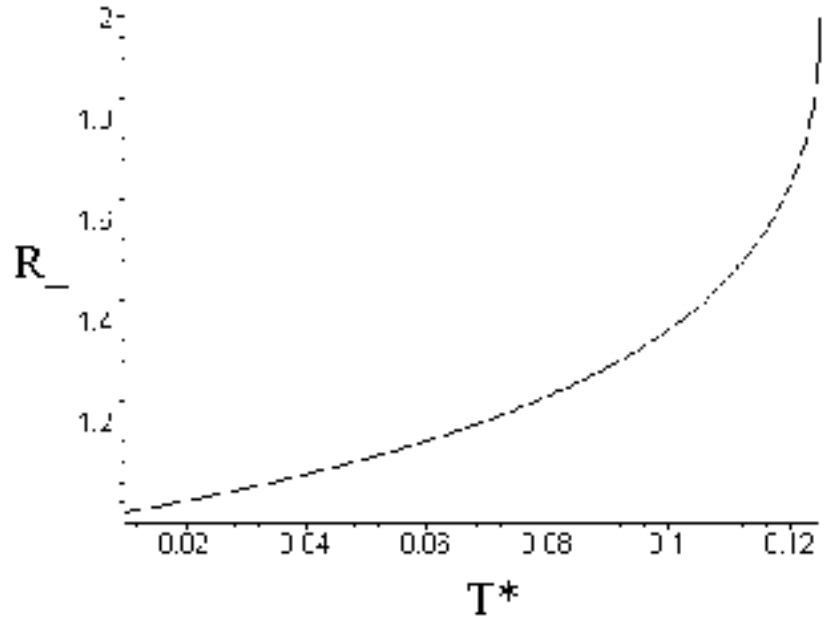}
\caption{The temperature dependence of $R_{-}$.} \label{R1}
\end{figure}

\begin{figure}
\includegraphics[scale=0.4]{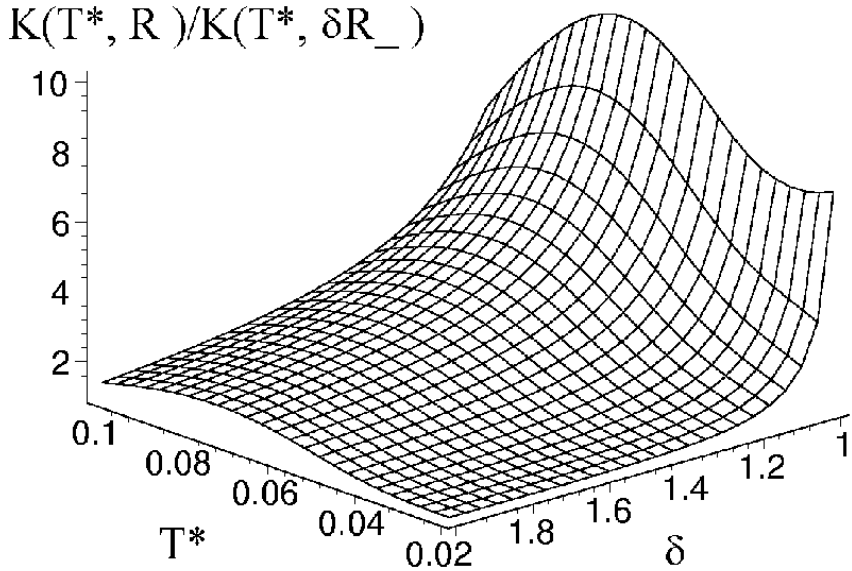}
\caption{The ratio $K(T^{*}, R_{+})/K(T^{*}, \delta  \cdot R_{-}
)$ as a function of $\delta$ and $T^*$.} \label{plateu}
\end{figure}

To define the size of a pair following  Bjerrum we investigate the
extremal points of the integrand in (\ref{as}). Doing so we get
two solutions:
\begin{equation}\label{rbj}
R_{-}(T^{*})=\frac{1-\sqrt{1-16T^{*}\lambda}}{4T^{*}}\, ,\quad
R_{+}(T^{*})=\frac{1+\sqrt{1-16T^{*}\lambda}}{4T^{*}}
\end{equation}
where
\[\lambda=\frac{L^{2}/2I_{0}}{q^2/a},\; I_{0}=\mu a^2.\]
Here $R_{+}$ is solution of the Bjerrum type (minimum of the
integrand in (\ref{as})), which as has been said above is
inappropriate. $R_{-}$ is another solution corresponding to the
maximum of the integrand, which has quite reasonable values and
correct behavior at low $T^{*}$. It is easy to check that
asymptotically for low values of the temperature $T^{*}$ the value
of $K(T^{*})$ is formed mainly by the maximum of the integrand. In
addition the appropriate limiting behavior to the hard-core
contact at formal limit $T^{*}\rightarrow 0$ is hold provided that
$\lambda=1/2$. This value of $\lambda$ is in full accordance with
the virial theorem \cite{land}. All these facts confirms that we
can treat the quantity $R_{-}$  as the size of the pair even at
"high" temperatures $T^{*}\leq 0.1$. In addition $R_{-}$ never
exceeds $2$, i.e. the interparticle the distance when the
influence of other pairs and charges on the effective potential
can be neglected (see Fig.~\ref{R1}).
So we put $R_{-}$ as the physical cutoff for (\ref{as}). The
existence of such length scale was noted in \cite{fishlev} basing
on the the numerical analysis of the function
$K(T^{*},R)/K(T^{*},R^{Bj})$, though only Coulomb potential was
included in Boltzmann factor. It gives the rate at which
$K(T^{*},R)$ rises very rapidly to its plateau value. In our case
we find the same behavior of $K(T^{*},R)$ at small temperatures,
$T^{*}<0.04$ (see Fig.~\ref{plateu}).
Finally we see that there is the natural temperature interval for
dipole fluid which is bounded from above by the temperature
\begin{equation}\label{tbj}
  T^{*}_{upper}\le\frac{1}{8}
\end{equation}
Therefore for $T  < T^{*}_{upper}$ the dipole is stable as itself.
Note that the existence of the temperature (\ref{tbj}) reminds
Kosterlitz-Thouless transition in 2D case. In particular, the
centrifugal energy introduced above plays the role analogous to
the chemical potential \emph{ "... required to create a pair of
particles of equal and opposite charge at a distance $r_{0}$ apart
..."} in Kosterlitz-Thouless model \cite{kt}. However, in contrast
to 2D case in 3D case there is no any divergence in the size of
the pair and therefore in its polarizability, but its derivative
on the temperature has singular behavior. In other words the
temperature derivative of the polarizability is singular but not
the polarizability itself. This inference might seem as mere an
artifact of introducing the upper cut off in (\ref{as}). But it
should be noted that taking dielectric permittivity $\epsilon$ as
the order parameter, which is directly connected with
polarizability, we get exactly the divergence of its temperature
derivative even in mean field approximation. This may serve as
additional support for the conjecture of intensive breaking of the
dipole pairs at the critical point observed in some numerical
experiments \cite{champ}.

The existence of the interaction between dipoles and the free
charges provides additional instability mechanism for their
dissociation thus reducing the temperature of "ideal" dissociation
(\ref{tbj}) because of the polarization of the dipole in the
external field of dipole-dipole potential (\ref{ddp}) and
Coulombic field of free charges. The consideration given above
states that there are two characteristic transition in the
dipole-dipole fluid: 1) "dipole liquid - dipole gas" critical
point of van der Waals type; 2) the smeared dissociation
"transition" from associated state to almost completely
dissociated one. This smeared transition can be characterized by
the temperature on the binodal at which the degree of dissociation
is 1/2.

The interaction between translational degrees of freedom of the
ions is characterized by the critical temperature of the
liquid-gas transition, while internal, rotational degrees of
freedom are involved into dissociation of such dipole fluid. These
degrees of freedom will strongly interact if the corresponding
potentials will be of the same magnitude i.e. $T_{c}\approx
T_{d}$. The additional confirmation of coincidence of such
transitions is the high degree of dissociation above the critical
point observed in numerical experiments \cite{g,champ} and
theoretical models \cite{gg}, which incorporate dielectric
permittivity resulting from the existence of the dipole pairs.
This means that $T_{d}$ cannot be less than $T_{c}$. All said
above means that in such situation we can not rely on (\ref{vdw})
since appropriate EOS should incorporate all relevant interactions
which lead to phase separation. In particular the critical
temperature is sensitive to the temperature dependence of the
parameter $A(T)$.

The condition for the dissociation of a pair in external
electrostatic field is:
\begin{equation}\label{disscond}
  <\mathbf{d}\cdot\mathbf{E}>= - <<U_{eff}>>\;,
\end{equation}
where
\begin{equation}\label{dip0}
  \mathbf{d}=\mathbf{d}_{0}+\mathbf{d}_{ind}
\end{equation}
is the dipole moment, which consists of proper and induced dipole
moments. Obviously,
\begin{equation}\label{disscond1}
  <\mathbf{d}_{0}\cdot\mathbf{E}>=0 \;, \quad
  <\mathbf{d}_{ind}\cdot\mathbf{E}>=\chi <\mathbf{E}^2>
\end{equation}
Here $\chi$ is the polarizability of a pair. The averaging over
the thermal equilibrium gives:
\begin{equation}\label{cav1}
<\mathbf{E}^2>=\frac{3k_{B}T}{\varkappa V_{ph}},
\end{equation}
where $\varkappa$ is the dielectric susceptibility of the medium
and $V_{ph}$ is the characteristic volume . It is connected with
the polarizability of the cavity. According to the definition:
\begin{equation}\label{cav2}
\varkappa=\frac{\epsilon-1}{4\pi}=\frac{1}{2}\chi\rho^*
\end{equation}
So we get the equation for the temperature in dimensionless form:
\begin{equation}\label{t}
\frac{6}{\rho^{*}V_{ph}}T^{*}=-<<U_{eff}>>
\end{equation}
Within the proposed approach we put $V_{ph}=\frac{4\pi}{3}
l_{c}^3$ where $l_{c}$ is the radius of first coordination sphere.
This is the minimal volume for which the conception of continuity
of the medium can be applied. By the order of magnitude
$l_{c}\,\approx\, 1.5\,a$. The solution of (\ref{t}) gives the
dependence $T^{*}(\delta)$. In order to fix the value of $\delta$
in the critical point which determine the size of the pair we
should equate $T^{*}(\delta)$ and $T^{*}_{c}(\delta)$ obtained
above. This way we get:
\begin{equation}
\label{c1} T^{*}_{c}= 0.048\;, \quad \rho^{*}_{c}= 0.054\,,\;\quad
P^{*}_{c}=4.8\cdot 10^{-4}\;,\quad \sigma=1.8\;,\quad  Z_{c}=0.19,
\end{equation}
which are close to those obtained above (\ref{critest}). In
notations of \cite{fishlev}, $\sigma = 2a_{2}$. In this work the
estimate for the parameter $a_{2}$ from simple geometric
considerations was given: $0.825 \le a_{2}\le 1.565$. Thus our
estimate is in this interval. From the results obtained above we
can infer that the dipole fluid of rotating dipoles in the
vicinity of its liquid-gas critical point is about to dissociate.
Sure our consideration is incomplete since it does not take into
account the existence of free charges.

Finally we estimate the Ginzburg number by the formula used for
the molecular liquids \cite{land}:
\begin{equation}  \label{gi}
Gi= \left(\frac{r_{0}}{\xi_{0}}\right)^6\,
\end{equation}
where $r_{0}= <<r>>\approx a$ is the interparticle spacing within
ionic pair and $\xi_{0}$ is the amplitude of the correlation
length for density fluctuations. Since the density fluctuations
are connected with the ones for dipole pairs we put it to be equal
$\xi\geq\sigma$. Using the parameters of the critical point found
in (\ref{c1}) we get the estimate:
\begin{equation}  \label{gi1}
Gi \le 0.04
\end{equation}

\section*{Discussion}

The ionic and dipole liquids form two natural approximations to
describe the critical properties of the systems similar the molten
NaCl. In our paper we have estimated the main critical parameters
for liquid with hard dipole as well as consider the influence of
the effects arising due to softness of a dipole molecule. In
particular the last is very important to describe the dielectric
properties of a system near the critical point. Besides, the
variation of molecule parameters due to the rotations allows us to
determine the equilibrium size of a ionic pair.

It is not excluded that the quantum corrections to internal states
of the dipole pairs will also slightly change the estimates. In
particular the temperature dependence of the vibrational
contributions to the heat capacity can also be studied. The
following step is to construct the equation of state for small
"soft" dipole molecules and to take into account the dissociation
process with the help of perturbation theory. The combination of
such an approach with that developed in \cite{kmt} on the basis of
ionic liquid allows to narrow the region of the most probable
values for the critical parameters.

Our estimate for the critical temperature correlates with the
known analytical results. Note that most of the analytical
approaches based on EOS for low density Coulombic system (DH, MSA
etc.) where the dissociation is taken into account perturbatively.

Within the dipole liquid approach we have obtained the estimate
for the Ginzburg temperature and have shown that it less than one
for the simple liquid by a factor $10^{-2} - 10^{-1}$. The
approximation of the dipole liquid allows us to analyze in the
evident form the contribution of the polarizational effects
\cite{kmt}. One can show that the lasts lead to the further
considerable decrease of the Ginzburg temperature.

Note also the possibility for the appearance of new inhomogeneous
phase near the critical point of ionic liquids. Since the
dissociation temperature $T_{d}$ is near $T_{c}$, the system can
desintegrate on the regions with the essentially different values
of the ionization degree $\Delta$: the drops of ionic and dipole
liquids. As a consequence the region with the Ising-like behavior
cannot be reached. This scenario needs in very careful
investigation. These and other questions will be the subjects of
further works.

\begin{acknowledgments}
The authors cordially thank to Professor V.M. Adamyan for fruitful
discussion of obtained results.
\end{acknowledgments}


\begin{thebibliography}{8}

\bibitem{mart} Ph.A. Martin, Rev. Mod. Phys. \textbf{60}, 1075 (1988)

\bibitem{dh} P. Debye and E. H\"uckel, Phyzik Z.  \textbf{24}, 195, 305 (1923)

\bibitem{bj} N. Bjerrum, Kgl. Dan. Vidensk. Selsk.
Mat.-Fys. Medd., \textbf{7}, 1 (1926)

\bibitem{msa} E. Waisman and J. Lebowitz, Journ. Chem. Phys. \textbf{56}, 3086
(1972); \textbf{56}, 3093 (1972)

\bibitem{fish}  M.E. Fisher, J. Stat. Phys. \textbf{75}, 1 (1994).

\bibitem{fishlev}  Y. Levin and M.E. Fisher, Physica A, \textbf{225}, 164
(1996).

\bibitem{st2}  G. Stell, J. Stat. Phys. \textbf{78}, 197 (1995).

\bibitem{gg}  B. Guillot and Y. Guissani, Mol. Phys. \textbf{87}, 37
(1996).

\bibitem{ws} Weing\"{a}rtner H. and Schr\"{o}er W., in {\it
Adv. Chem. Phys.}, edited by I. Prigogine and S.A. Rice (Wiley,
London, 2001), Vol. 116, p. 1.

\bibitem{champ}  P.J. Camp and G.N. Patey, Phys. Rev. E, \textbf{60}, 1063
(1999).

\bibitem{mc} J.M. Caillol, D. Levesque and J.J. Weis, J.
Chem. Phys., \textbf{107},1565, (1997).

\bibitem{gr} G. Orkoulas and A.Z. Panagiotopoulas, J.
Chem. Phys., \textbf{110},1581, (1999).

\bibitem{dip}  O.A. Osipov,  V.I. Minkin and A.D. Garnovskii
(eds.), {\it Spravochnik po dipol'nym momentam veschestv} (Moscow,
Vysshaya Shkola, 1971).

\bibitem{kt} J. M. Kosterlitz and D. J. Thouless,
J. Phys. C, \textbf{6}, 1181 (1973).

\bibitem{land} L. D. Landau and E. M. Lifshits, {\it Statistical Physics} (Pergamon,
Oxford, 1980).

\bibitem{g}  M.J. Gillan, Mol. Phys., \textbf{41}, 75
(1980).

\bibitem{kmt}  V.L. Koulinskii, N.P. Malomuzh and V.A. Tolpekin, Phys. Rev.
E, \textbf{60}, 6897 (1999).

\end{thebibliography}
\end{document}